\documentclass[final, 3p, 12pt]{elsarticle}

\usepackage{amssymb}
\usepackage{amsmath}
\usepackage{amsthm}

\usepackage{float}
\usepackage{longtable}
\journal{Journal of Complexity}
\newtheorem{thm}{Theorem}
\newtheorem{algor}[thm]{Algorithm}

\newtheorem{lem}[thm]{Lemma}
\newtheorem{defn}[thm]{Definition}

\begin{document}
\newcommand{\etal}[1]{\emph{et al.}~#1}
\newcommand{\oankm}{\text{OA}(N,k-1,s,t)}
\newcommand{\oan}{\text{OA}(N,k,s,t)}\newcommand{\oano}{\text{OA}(N,k_0,s,t)}
\newcommand{\oank}{\text{OA}(N,k+1,s,t)}
\newcommand{\oal}{\text{OA}(\lambda s^t,k,s,t)}
\newcommand{\fnst}{f(N,s,t)}
\newcommand{\fnt}{f(N,2,t)}
\newcommand{\ftnt}{f(2N,2,t+1)}
\newcommand{\fcan}{f^{\text{CA}}_{\lambda}(N,s,t)}
\newcommand{\fpan}{f^{\text{PA}}_{\lambda}(N,s,t)}
\newcommand{\fcant}{f^{\text{CA}}_{\lambda}(N,2,t)}
\newcommand{\fpant}{f^{\text{PA}}_{\lambda}(N,2,t)}

\newcommand{\kin}{k_{\mathrm{input}}}
\newcommand{\kout}{k_{\mathrm{out}}}
\newcommand{\oatntmm}{\text{OA}(2N,k,2,t+1)}
\newcommand{\oatntm}{\text{OA}(2N,k+1,2,t+1)}
\newcommand{\oantf}{\text{OA}(2N,k+1,2,t+1)}
\newcommand{\oantff}{\text{OA}(2N,k+2,2,t+1)}
\newcommand{\oanst}{\text{OA}(N,k,s,2)}
\newcommand{\oantm}{\text{OA}(N,k-1,2,t)}
\newcommand{\oant}{\text{OA}(N,k,2,t)}
\newcommand{\oantt}{\text{OA}(N,k+1,2,t)}
\newcommand{\oalt}{\text{OA}(\lambda 2^t,k,2,t)}
\newcommand{\fxoptt}{f(\X^{opt}_{N})}
\newcommand{\oam}{\text{OA}(N,s^{k_1}_1s^{k_2}_2\cdots s^{k_v},t)}
\newcommand{\yoptt}{\Y^{opt}_{N,k}}
\newcommand{\xoptt}{\X^{opt}_{N,k}}
\newcommand{\x}{\boldsymbol{x}}
\newcommand{\M}{\boldsymbol{M}}
\newcommand{\N}{\boldsymbol{N}}
\newcommand{\HH}{\boldsymbol{H}}
\newcommand{\I}{\boldsymbol{I}}
\newcommand{\1}{\boldsymbol{1}}
\newcommand{\zz}{\boldsymbol{0}}
\newcommand{\aaa}{\boldsymbol{a}}
\newcommand{\bb}{\boldsymbol{b}}
\newcommand{\cc}{\boldsymbol{c}}
\newcommand{\uu}{\boldsymbol{u}}
\newcommand{\rrr}{\boldsymbol{r}}
\newcommand{\xx}{\boldsymbol{x}}
\newcommand{\y}{\boldsymbol{y}}
\newcommand{\dd}{\boldsymbol{d}}
\newcommand{\vvec}{\boldsymbol{vec}}
\newcommand{\Y}{\boldsymbol{Y}}
\newcommand{\X}{\boldsymbol{X}}
\newcommand{\D}{\boldsymbol{D}}
\newcommand{\A}{\boldsymbol{A}}\newcommand{\tA}{\tilde{\boldsymbol{A}}}
\newcommand{\XX}{\boldsymbol{X}^{\top}\boldsymbol{X}}
\newcommand{\XT}{\boldsymbol{X}^{\top}}
\newcommand{\bs}{\boldsymbol}
\newcommand{\orr}{\text{\ or \ }}
\newcommand{\md}[2][4]{#2\ \ (\mathrm{mod}\ \ #1)}
\newcommand{\rr}{\mbox{I\!\!\,R}}
\newcommand{\es}{\mathrm{E}(s^2)}
\newcommand{\smax}{s_{\mathrm{max}}}
\newcommand{\freq}{f_{\smax}}
\newcommand{\dfk}[1]{\Delta f^k(#1)}
\newcommand{\Z}{\mathbb{Z}}
\newcommand{\mth}[1]{\multicolumn{3}{c|}{#1}}
\newcommand{\mtw}[1]{\multicolumn{2}{c|}{#1}}
\newcommand{\GWP}{\mathrm{GWP}}
\begin{frontmatter}
\title{Algorithms for finding generalized minimum aberration designs}
\author[AFIT]{Dursun A.~Bulutoglu\corref{cor1}}
\ead{Dursun.Bulutoglu@afit.edu}
\author[WVU]{Kenneth J.~Ryan}
\ead{kjryan@mail.wvu.edu}
\address[AFIT]{Department of Mathematics and Statistics, Air Force Institute of Technology,\\Wright-Patterson Air Force Base, Ohio 45433, USA}
\address[WVU]{Department of Statistics, West Virginia University,\\ Morgantown, West Virginia 26506, USA}
\cortext[cor1]{Corresponding author}
\begin{abstract}
Statistical design of experiments is widely used in scientific and industrial investigations.  A generalized minimum aberration (GMA) orthogonal array is optimum under the well-established, so-called GMA criterion, and such an array can extract as much information as possible at a fixed cost. Finding GMA arrays is an open (yet fundamental) problem in design of experiments because constructing such arrays becomes intractable as the number of runs and factors increase.
We develop two directed enumeration algorithms that call the integer programming with isomorphism pruning algorithm of Margot~\cite{Margot07} for the purpose of finding GMA arrays. Our results include 16 GMA arrays that were not previously in the literature, along with documentation of the efficiencies that made the required calculations possible within a reasonable budget of computer time. We also validate heuristic algorithms against a GMA array catalog, by showing that they quickly output near GMA arrays, and then use the heuristics to find near GMA arrays when enumeration is computationally burdensome.
\end{abstract}

\begin{keyword}
Constraint programming; Directed enumeration constraints; Extension algorithm; Heuristic search; Isomorphism rejection; \texttt{Nauty}
\end{keyword}
\end{frontmatter}
\section{Introduction}
This work tailors some state-of-the-art methods from operations research to find solutions in a fundamental class of problems in design of experiments. The main contribution of this paper is two directed enumeration algorithms that call the Margot~\cite{Margot07} integer linear programming (ILP) solver. These algorithms were used to extend the known catalog of optimum orthogonal arrays (OAs) with 16 new entries. We also use heuristic search algorithms for finding optimum or near-optimum OAs when exact methods require too much computation. Algorithm performance (i.e., speed and ability to find optimum solutions) is documented.
\subsection{Orthogonal arrays and the GMA criterion}
A factorial design $\Y$ with $N$ runs and $k$ factors
each having $s$-levels is an orthogonal array of strength $t$, $1
\leq t \leq k$, denoted by $\oan$, if each of the $s^t$ level
combinations appears exactly $N/s^t$ times when $\Y$ is projected
onto any $t$ factors. The index $\lambda$ of an $\oan$  is defined
as $N/s^t$. An $\oan$ is universally optimal for estimating the
model containing all main effects and all interactions having
$\lfloor t/2 \rfloor$ factors or less; see Cheng~\cite{Cheng}  and
Mukerjee~\cite{Mukerjee}.

The design obtained by permuting factors or runs as well as levels
in a subset of factors in an $\oan$ is also an $\oan$. Let such
operations be called {\em isomorphism operations}.  Two $\oan$ are
called {\em isomorphic} if one can be obtained from the other by
applying a sequence of isomorphism operations. Assuming the
hierarchical ordering principle (see Section 3.5 of Wu and Hamada~\cite{Wu}), two $\oan$ are compared under model uncertainty using the
{\it generalized minimum aberration} (GMA) criterion developed in Xu and
Wu~\cite{Xu01}. Let $\Y=[y_{ij}]$ be a $2$-level design with entries
$\pm 1$ having $N$ runs and $k$ factors, and let
$l=\{i_1,i_2,\ldots,i_r\}\subseteq\Z_k :=\{1,\ldots,k\}$ be a
nonempty subset of $r$ factors.  The GMA criterion is based on the
concept of the $J$-characteristics
\begin{equation*}
J_r(l):=\sum_{i=1}^N{\prod_{j \in l}{y_{ij}}}
\end{equation*}
of Tang and Deng~\cite{Tang}. Note $0\leq |J_r(l)|\leq N$, and a larger $|J_r(l)|$ implies a stronger degree of aliasing among the factors in $l$. An average aliasing among all subsets of $r$ factors is
\begin{equation*}
A_r(\Y):=\frac{1}{N^2}\sum_{\{l \subseteq \Z_k:|l|=r\}}J_r(l)^2,
\end{equation*}
and $\GWP(\Y):=(A_1(\Y),A_2(\Y),\ldots,A_k(\Y))$ is the {\it generalized word length pattern} (GWP) of $\Y$. The GMA criterion selects designs that sequentially minimize the GWP. A design with the same first non-zero GWP entry as a GMA design is a {\it weak} GMA design.

The general concept of GWP for $s$-level designs is computed as
follows. Let $d_{ij}(\Y)$ be the number of columns at which the
$i$th and $j$th rows of $\Y$ differ, and define
\begin{equation*}
B_r(\Y):=N^{-1}|\{(i,j):d_{ij}(\Y)=r, \ i,j=1,\ldots,N\}|
\end{equation*}
for $r=0,\ldots, k$. The {\it distance distribution}
($B_0(\Y),B_1(\Y),\ldots,B_k(\Y)$) of $\Y$ determines the GWP and vice versa; the direct relationships provided in Xu and
Wu~\cite{Xu01} are:
\begin{eqnarray}
A_j(\Y) &=& N^{-1}\sum_{i=0}^k P_j(i,s,k)B_i(\Y) \nonumber \\
B_j(\Y) &=& Ns^{-k}\sum_{i=0}^{k}P_j(i,s,k)A_i(\Y) \nonumber
\end{eqnarray}
for $j=0,\ldots,k$, where $A_0(\Y)=1$ and $P_j(x,s,k) :=
\sum_{i=0}^{j}{(-1)^i(s-1)^{j-i}\binom{x}{i}\binom{k-x}{j-i}}$ are
the {\em Krawtchouk polynomials}.  When computing the Krawtchouk
polynomials, the recursion
\begin{equation*}
P_j(x,s,k)=P_j(x-1,s,k)-P_{j-1}(x-1,s,k)-(s-1)P_{j-1}(x,s,k)
\end{equation*}
with initial values $P_0(x,s,k)=1$ and
$P_j(0,s,k)=(s-1)^j\binom{k}{j}$ is useful.

\subsection{Finding GMA designs}\label{sec:finding}
In general, finding GMA designs is a very difficult problem.
Butler~\cite{Butler03,Butler04} theoretically constructed 2-level GMA designs.
Butler's proofs for establishing that the constructed designs were
GMA involved finding lower bounds for the GWP of 2-level designs
for a certain number of runs and factors. Xu~\cite{Xu05} derived lower
bounds for the GWP using linear programming in infinite precision
and also found 2-level factorial designs based on the
Nordstorm-Robinson code that achieve the bounds. Bulutoglu and
Kaziska~\cite{BulutogluK} improved the lower bounds of Xu~\cite{Xu05}  using ILP
in infinite precision developed by Espinoza~\cite{Espinoza} and Applegate \etal{\cite{Applegate}}.
 Fang, Zhang, and Li~\cite{Fang} and Sun, Liu, and Hao~\cite{Sun} developed lower bounds and algorithms for finding GMA designs. However, these algorithms are not guaranteed to return a GMA design, and a GMA design can be identified only if the best design found achieves the sharpest known GWP lower bound.

Another way of finding GMA designs is by classifying all
non-isomorphic OAs. If two $\oan$ are isomorphic, they are
indistinguishable under the GMA criterion. On the other hand,
there are non-isomorphic $\oan$ with the same GWP. Classifying all
non-isomorphic $\oan$ allows the best to be found with respect to
the GMA or any other ordering criterion that is invariant
between isomorphic designs. Many have studied the problem of
classifying all non-isomorphic $\oan$ (e.g., Stufken and Tang~\cite{Stufken}; Bulutoglu and Margot~\cite{BulutogluM}; Schoen, Eendebak, and Nguyen~\cite{Schoen}; and Bulutoglu and Ryan~\cite{BulutogluR}).

Bulutoglu and Margot~\cite{BulutogluM} showed that finding all $\oan$ is
equivalent to finding all nonnegative integer solutions to a
symmetric ILP with binary coefficients. Two solutions are defined
to be {\em isomorphic} if they correspond to isomorphic OAs, so
finding all non-isomorphic solutions is equivalent to finding all
non-isomorphic $\oan$. Branch-and-cut algorithms (see Padberg and
Rinaldi,~\cite{Padberg}) are a standard technique for solving ILPs, but a
presence of symmetry in an ILP  requires extending the basic
branch-and-cut algorithm to avoid solving isomorphic
subproblems. Such an extension developed by Margot~\cite{Margot02,Margot03a,Margot03b,Margot07} was used by Bulutoglu and Margot~\cite{BulutogluM} to find all
isomorphism classes of $\oan$ for many $(N,k,s,t)$ combinations.


Schoen, Eendebak, and Nguyen~\cite{Schoen} developed the Minimum Complete Set (MCS) algorithm for enumerating OAs up to isomorphism. MCS is a constraint programming (CP) algorithm with isomorphism rejection. Schoen, Eendebak, and Nguyen~\cite{Schoen} used their MCS algorithm to find all non-isomorphic $\oan$ for many $(N,k,s,t)$ combinations--including all those in Bulutoglu and Margot~\cite{BulutogluM}--and mixed level OAs where all factors do not have the same number of levels. Bulutoglu and Ryan~\cite{BulutogluR} introduced orthogonal design equivalence and four algorithms based on ILP and ILP with isomorphism pruning developed by Margot~\cite{Margot07}. This reduced the necessary computational burden and enabled them to find all non-isomorphic OA$(160,9,2,4)$, OA$(160,10,2,4)$, OA$(176,8,2,4)$, OA$(176,9,2,4)$, and OA$(176,10,2,4)$ as well as the GMA arrays for these $(N,k,s,t)$ combinations.

The main contribution of this work is two new directed enumeration algorithms, which capture all $\oant$ that are as good or better than a pre-specified GWP. Directed enumeration requires the algorithms of Bulutoglu and Ryan~\cite{BulutogluR}, so directed enumeration is defined in Section \ref{sec:direct} after a brief overview of Bulutoglu and Ryan~\cite{BulutogluR} in Section \ref{sec:enum}. A theoretical justification for directed enumeration is in \ref{sec:proofs}, and results include the 16 newly established 2-level GMA designs described in Section \ref{sec:results}. The actual designs are available upon request. Another contribution is the heuristic search algorithms in Section \ref{sec:heuristic}, similar to those of Fang, Zhang, and Li~\cite{Fang}. Heuristics are validated against the extensive GMA design catalog in \ref{sec:cat}, by demonstrating that they quickly output GMA or near GMA $\oan$. The heuristics are also used to output $12$ weak GMA OA($36,k,2,2$) and GMA or near GMA OA($28,k,2,2$) for large $k$, when enumerations take too long. The extensive GMA catalog in \ref{sec:cat} provides the distance distribution for each entry and is a valuable resource for other researchers.  Computations were all done at the Ohio Supercomputer Center with 2.6 GHz processors or on the second author's computer with 3GHz processors.
\section{Extension algorithms for enumerating {\boldmath  OA$(N,k,$ $s,t)$}}\label{sec:enum}
Necessary results and algorithms from Bulutoglu and Ryan~\cite{BulutogluR} are restated for convenience. Let $f(N,s,t)$ be the largest $k$ such that an $\oan$ exists. For $k=t+1,\ldots,f(N,s,t)+1$, let $n_k$ be the number of non-isomorphic $\oan$ and $T_k=\{\Y_1, \ldots, \Y_{n_k}\}$ be a set of non-isomorphic $\oan$. A Generic Extension Algorithm \ref{alg:basic} enumerates $T_k$ from $T_{k-1}$.
\begin{algor}[Generic extension, Bulutoglu and Ryan \cite{BulutogluR}] \label{alg:basic}  Input: $N$, $k$, $s$, $t$,
$T_{k-1}$, $l:=1$.
\begin{enumerate}

\item Obtain a set $M_l$ of $\oan$ such that the first $k-1$ columns are $\Y_l \in T_{k-1}$ and exactly 1 representative from each isomorphism class of such $\oan$ is included. This can be done either by some CP- or ILP-based method.\label{item:extend}

\item Increment $l:=l+1$ and then repeat Step \ref{item:extend} if $l \leq n_{k-1}$.

\item \label{item:M} Set $M:=\bigcup_{l=1}^{n_{k-1}}M_l$. Form $T_k$ from $M$ by picking one representative from each isomorphism class of $\oan$ in $M$. Output: $T_k$.\label{item:nauty}

\end{enumerate}
\end{algor}
The $s^t$ full factorial design replicated $\lambda:=N/s^t$ times is the singleton $T_t$.  Given $T_{k-1}$, $T_{k+m}$ is obtained after applying an extension algorithm $m+1$ times. Alternatively, if $k>t+1$, Bulutoglu and Margot~\cite{BulutogluM} can be used to directly obtain input $T_{k-1}$ to an extension algorithm. Throughout this and our previous work, we always used the graph-based approach with the program \texttt{nauty} (McKay~\cite{McKay13nauty}) to execute Step \ref{item:nauty} (McKay and Piperno~\cite{McKay13}; Ryan and Bulutoglu~\cite{Ryan}). Algorithmic contributions of Bulutoglu and Ryan~\cite{BulutogluR} included computationally efficient versions of Step \ref{item:extend}. Let the factor levels of an $\oan$ be coded $0,\ldots,s-1$. Lemma \ref{lem:Y'} and Definition \ref{def:cor} from Bulutoglu and Ryan~\cite{BulutogluR} were used to formulate the ILP feasibility problem for extending an $\oankm$ to an $\oan$. By Lemma \ref{lem:Y'} and Definition \ref{def:cor}, Algorithm \ref{alg:BR3idd} follows as a specific version of Algorithm \ref{alg:basic}.
\begin{lem}\label{lem:Y'}
Let $\Y$ be an $N$ run, $k$ factor, $s$-level factorial design with
columns $\{\y_1,\y_2,\ldots,$ $\y_k\}$. Let $\Y'$ be the $N \times
(s-1)k$ matrix with columns $\{\y'_1,\y'_2,\ldots,\y'_{(s-1)k}\}$ where for
$j=1, \ldots, k$ and $r=1, \ldots, s-1$ the $i$th entry of
$\y'_{(s-1)(j-1)+r}$ is $1$ if the $i$th entry of $\y_j$ is $r-1$ and is
$0$ otherwise.  Then $\Y$ is an $\oan$ if and only if for $q=1,\ldots,t$
\begin{equation}\label{eqn:jy'01}
\sum_{i=1}^Ny'_{ih_1}y'_{ih_2}\cdots\,y'_{ih_q}=\frac{N}{s^q} \\
\end{equation}
for any $q$ columns $\{\y'_{h_1},\y'_{h_2},\ldots,\y'_{h_q}\}$ of
$\Y'$ such that $\lceil h_{i'}/(s-1) \rceil \ne \lceil h_{j'}/(s-1) \rceil$ for
all $1 \leq i'<j' \leq (s-1)k$.
\end{lem}
\begin{defn}\label{def:cor}
Let $\Y'$ as in Lemma \ref{lem:Y'} be a solution to the system of Equations (\ref{eqn:jy'01}). Also, let $\Y$ be the OA$(N,k,s,t)$ obtained from $\Y'$ with columns
\begin{equation*}
\Y_{j}=\sum_{r=1}^{s-1}{(r-1)\y'_{(s-1)(j-1)+r}}+(s-1)\left(\1_N-\sum_{r=1}^{s-1}{\y'_{(s-1)(j-1)+r}}\right)\quad \text{for each}\quad j=1,\ldots,k,
\end{equation*}
where $\1_N$ is an $N \times 1$ vector of $1$s.
Then $\Y$ is called {\em the $\oan$ corresponding to $\Y'$}.
\end{defn}
\begin{algor}[Identity group, Bulutoglu and Ryan~\cite{BulutogluR}] \label{alg:BR3idd}
Enumerate all solutions to the following ILP in Step \ref{item:extend} of Algorithm \ref{alg:basic} to enumerate the corresponding OA extensions.
\begin{enumerate}
\item[] Construct $\Y'_l$ from $\Y_l$ using
    Lemma \ref{lem:Y'}, extend $\Y'_l$ with $s-1$ columns of binary
    variables $\xx_r=(x_{1r},x_{2r},\ldots,x_{Nr})^\prime$ for
    $r=1,\ldots,s-1$, and obtain all solutions to the ILP:
\begin{align}\label{ilp:oa}
&\quad \quad \quad \quad   \quad \min  \quad 0 \nonumber\\
& \text{subject to:} \quad \sum_{i=1}^Nx_{ir}=\frac{N}{s}, \nonumber \\
&\sum_{i=1}^Ny'_{ih_1}y'_{ih_2}\cdots\,y'_{ih_{q-1}}x_{ir} =\frac{N}{s^q},\\
&x_{1,1}=1, \ \ \ \sum_{r=1}^{s-1}x_{ir} \leq 1, \ \ \ \sum_{m=1}^{r}{(x_{i''m}-x_{j''m})}
\ge 0, \nonumber\\
 &  \xx_r \in \{0,1\}^{N}, \quad i=1,\ldots,N, \quad r=1,\ldots,s-1, \quad q=2,\ldots,t \nonumber
\end{align}
 for each pair of equal rows in the input design with indices $1 \le i''< j'' \le N$ and for any $q-1$ columns $\y'_{h_1}, \y'_{h_2},\ldots\,\y'_{h_{q-1}}$ as in Lemma \ref{lem:Y'} with the last $(s-1)$ columns of $\Y'$  deleted.  Each solution matrix $[\xx_1,\x_2,\ldots,\x_{s-1}]$ to ILP (\ref{ilp:oa}) has a corresponding $\oan$. Take $M_l$ to be the set of all such $\oan$.
\end{enumerate}
\end{algor}

By Lemma \ref{lem:Y'}, the extension problem is that of constraint feasibility. The constant objective function of 0 in ILP (\ref{ilp:oa}) was simply a convenience, so that existing ILP solvers could be used directly. It was mentioned previously in Section \ref{sec:finding} that symmetry in ILP (\ref{ilp:oa}) results in redundant subproblem nodes in the enumeration tree. Incorporating isomorphism pruning as described next can boost algorithm speed by removing redundant nodes. For an $\oan=\Y$, let $\Y''$ be the $N\times sk$ matrix obtained from $\Y'$ in Lemma \ref{lem:Y'} by concatenating columns $\1_N-\sum_{r=1}^{s-1}{\y'_{(s-1)(j-1)+r}}$ for $j=1,\ldots,k$. Let $$G_s^k=\{\pi \in S_N| \mbox{ there exists } \sigma \in S_{s^k} \mbox{ such that } \Y''(\pi,\sigma)=\Y''\},$$ where $S_m$ is the group of all permutations of $m$ objects and $\Y''(\pi,\sigma)$ is the resulting matrix when the rows and columns of $\Y''$ are permuted according to $\pi$ and $\sigma$, respectively. Let $H_s^{k}$ be the maximum size subgroup of $G_s^k$ that does not have an element that sends a row's index of $\Y$ to the index of an equal row. Algorithm \ref{alg:BR3idd} can be run using  the group $H_s^{k-1}$ with the isomorphism pruning of Margot~\cite{Margot07} in a branch-and-cut algorithm to implement isomorphism rejection while solving the needed ILPs.

With larger $N$ and moderate $k$, using isomorphism pruning in Algorithm \ref{alg:BR3idd} is vastly superior. For example,
extending OA(24,5,2,2) to OA(24,6,2,2) takes 1.52 and 4.45 minutes with and without isomorphism pruning. However,  even with large $N$, not using isomorphism pruning in Algorithm \ref{alg:BR3idd}  is most efficient with large $k$ because for such cases  $|H_{s}^{k-1}|=1$ and isomorphism pruning is a redundant computational overhead (Bulutoglu and Ryan~\cite{BulutogluR}).

\section{Directed enumeration algorithms for finding GMA {\boldmath $\oant$}}\label{sec:direct}
Stufken and Tang~\cite{Stufken} used $J$-characteristics to classify all OA($N,t+2,2,t$). The following generalization of their Lemma 3 was proved by Bulutoglu and Kaziska~\cite{BulutogluK} and is used in this work to find GMA designs when full enumerations from Section \ref{sec:enum} are computationally intractable due to large $|T_k|$.
\begin{lem}\label{lem:Stufken}
For an $\oant$ with $N=\lambda 2^t$ and $k \geq t+2$ the following hold.
\begin{enumerate}

\item For any $l\subseteq \{1,\ldots,k\}$, $J_{|l|}(l)=\mu_l2^t$ for some integer $\mu_l$.

\item If $\lambda$ is even, then $\mu_l$ is even.

\item  If $\lambda$ is odd, then $\mu_l$ is odd if $\binom {|l|-1}{|l|-t-1} \equiv 1$  (mod $2$) and even otherwise.

\end{enumerate}
\end{lem}
Let $J_{t+1}(l)$ be a $J$-characteristic of an $\oant$ based on some
subset $l$ of $t+1$ factors. By Lemma \ref{lem:Stufken},
$J_{t+1}(l)=\mu_{l} 2^t$ for some integer $\mu_{l}$, and $\mu_{l}$
is odd if and only if $\lambda$ is odd. Now, the set of all $\oant$
such that all $|J_{t+1}(l)|$ achieve their lower bound of $2^t$ with
odd $\lambda$ will contain all GMA designs if this set is nonempty;
let $T_k^\prime$ be this subset of $T_k$.  Algorithm \ref{alg:direct1} is intended
 for directed enumerations when $\lambda$ is odd. One can increase $t$ by $1$
and use an algorithm from Section \ref{sec:enum} when $\lambda$ is even.

\begin{algor}[Minimizing all $|J_{t+1}(l)|$] \label{alg:direct1}
Use Algorithm \ref{alg:BR3idd} after replacing $T_{k-1}$ and $T_k$ with
$T_{k-1}^\prime$ and  $T_k^\prime$ and adding the
2 directed enumeration constraints $(\lambda-1)/2 \le
\sum_{i=1}^{\lambda}{x_{j_i1}} \le (\lambda+1)/2$ to ILP
(\ref{ilp:oa}) for every subset of $\lambda$ runs $j_i$, i.e., $1
\le j_1 < \cdots < j_{\lambda} \le N$, whose elements replicate a run in the
$2^t$ full factorial design when the input design is projected
onto $t$ columns. (There are $2\binom{k-1}{t}2^t$ new directed
enumeration constraints, but still use group $H_2^{k-1}$ from Algorithm \ref{alg:BR3idd} if
isomorphism pruning.)
\end{algor}

Lemmas \ref{lem:divisibility} and \ref{lem:recursion} can be used to further direct/constrain the enumeration to a smaller subset of good designs which includes GMA designs if nonempty. Lemma \ref{lem:divisibility} follows from Lemma \ref{lem:Stufken} and is used in Lemma \ref{lem:recursion}. Lemma \ref{lem:recursion} generalizes Lemma 2 in Xu~\cite{Xu09} and justifies the more stringent directed enumeration given by Algorithm \ref{alg:direct2}.

\begin{lem}\label{lem:divisibility}
Let $\Y$ be an $\oant$. Then
$[N^2A_j(\Y)-\phi_1(k,t,j)]/\phi_2(t,j)$ are nonnegative integers
for $j=t+1,\ldots,k$, where $\phi_1(k,t,j) :=
I(t,j)\binom{k}{j}2^{2t}$, $\phi_2(t,j) := 2^{2t+2+I(t,j)}$, and
$I(t,j):=1$ if $\binom {|l|-1}{|l|-t-1} \equiv 1$  (mod $2$) and
$I(t,j):=0$ otherwise.
\end{lem}
\begin{lem}\label{lem:recursion}
Let $\Y$ be an $\oant$ and $r < k$ be a positive integer. If
$\Y(-i)$ (i.e., design $\Y$ with column $i$ deleted) is GMA
among all other delete-one-factor projections of $\Y$, then
\begin{equation}\label{eqn:recursive}
A_{t+r}(\Y(-i))\leq \left \lfloor
\frac{N^2(A_{t+r}(\Y)-(t+r)A_{t+r}(\Y)/k)-\phi_1(k-1,t,t+r)}{\phi_2(t,t+r)}
\right
 \rfloor /N^2.
\end{equation}
\end{lem}
\begin{algor}[Minimizing all $|J_{t+1}(l)|$, and Recursion
(\ref{eqn:recursive}) on $A_{t+2}(\Y)$] \label{alg:direct2} Let
$T_k^{\prime\prime}$ be the subset of $T_k^\prime$ satisfying
Recursion (\ref{eqn:recursive}), with initial values $\kin \ge k$
and $A_{t+2}$. Replace each ``$\prime$" with ``$\prime\prime$" in
Algorithm \ref{alg:direct1} and discard designs not satisfying
Recursion (\ref{eqn:recursive}) after solving each ILP.
\end{algor}
\vspace{-.5cm}
\begin{table}[h]\scriptsize
\caption{Extending OA(20,$k-1$,2,2) to OA(20,$k$,2,2) with real times (minutes:seconds).}
\label{tab:compare} \centering \vspace{.1in}
\begin{tabular}{|c|rrc|rrc|rrc|}\hline
&\multicolumn{3}{|c|}{Algorithm \ref{alg:BR3idd}}
&\multicolumn{3}{|c|}{Algorithm \ref{alg:direct1}}
&\multicolumn{3}{|c|}{Algorithm \ref{alg:direct2}}\\
$k$& $|T_k|$&    $|M|$ &      Time  & $|T_k^\prime|$ & $|M^\prime|$ & Time  & $|T_k^{\prime\prime}|$ & $|M^{\prime\prime}|$ & Time  \\ \hline
 3 &      3 &        5 &   0:00.10           &1 &      4 &   0:0.16       &1 &      4 &   0:0.18     \\
 4 &      3 &       44 &   0:00.38          &2 &     14 &   0:0.19       & 1 &      8 &   0:0.19   \\
 5 &     11 &      363 &   0:02.06        &4 &     24 &   0:0.37         & 2 &      8 &   0:0.22 \\
 6 &     75 &     1659 &   0:09.30       &13 &    236 &   0:1.84       & 2 &     14 &   0:1.10 \\
 7 &    474 &     5679 &   0:38.37      &21 &    262 &   0:4.52        & 2 &     12 &   0:0.49 \\
 8 &   1603 &    15219 &   2:14.77    & 6 &     40 &   0:4.06           &0 &      0 &   0:0.32\\
 9 &   2477 &    22744 &   4:11.31     &2 &      6 &   0:0.92          &&&\\
10 &   2389 &    23984 &   4:56.12    &1 &      4 &   0:0.37          &&&\\
11 &   1914 &    21149 &   5:33.74    &0 &      0 &   0:0.17           &&&\\
12 &   1300 &    15272 &   4:19.80     &&&&&& \\
13 &    730 &     9100 &   2:48.59       &&&&&& \\
14 &    328 &     4380 &   1:31.58       &&&&&& \\
15 &    124 &     1640 &   0:38.18       &&&&&& \\
16 &     40 &      496 &   0:12.26         &&&&&& \\
17 &     11 &      120 &   0:03.38          &&&&&&\\
18 &      6 &       22 &   0:00.64           &&&&&& \\
19 &      3 &        6 &   0:00.27             &&&&&&\\ \hline
\end{tabular}
\end{table}

Enumerations of  OA(20,$k$,2,2) are used as a ``proof of concept" for directed enumerations, since both full and directed enumerations are quick. See Table \ref{tab:compare}. An efficiency of Algorithm \ref{alg:direct1} for small $k$ is the speed increase due to processing fewer designs; compare $|T_k|$ and $|T_k^\prime|$. If $|T_k^\prime| \ne 0$, the output $T_k^\prime$ from Algorithm \ref{alg:direct1} is the set of all weak GMA designs. Shortcomings of Algorithm \ref{alg:direct1} are a crude lower bound of $10 < 19=f(20,2,2)$ and the inability to capture GMA designs for large $k=11,\ldots,19=f(20,2,2)$. Similarly, note another speed increase due to further reduction in the number of designs because of Recursion (\ref{eqn:recursive}) with $A_4 \le 13.75$ and $\kin=10$; compare $|T_k^{\prime}|$ and $|T_k^{\prime\prime}|$. This boost in speed comes at the expense of an even cruder lower bound of $7 < f(20,2,2)$. Assume odd $\lambda$.  Algorithm \ref{alg:direct1} can be used for given $(N,s,t)$ to find GMA designs for small and medium $k$ when the full enumeration methods of Section \ref{sec:enum} are computationally too intense.  Algorithm \ref{alg:direct2} can be used if Algorithm \ref{alg:direct1} requires too much computation, but then only smaller $k$ GMA designs will be obtained. This approach was used to further extend the catalog of GMA designs, when $N=28,36$.

\section{Extending the catalog of GMA {\boldmath $\oan$}}\label{sec:results}
Directed enumerations of 28 and 36 run cases are summarized in Table \ref{tab:enum}. When $N=28$, Algorithm \ref{alg:direct1} was used, so the number of classes reported is $|T_k^\prime|$. When $N=36$, Algorithm \ref{alg:direct2} was used, so the number of classes reported is $|T_k^{\prime\prime}|$, where the initial values for Recursion (\ref{eqn:recursive}) were $\kin=18$ and $A_4=120,000/36^2$. Initial values for Recursion (\ref{eqn:recursive}) were determined based on combining the values $A_4$ of near GMA OA$(N,\kin,2,t)$ obtained by heuristic search from Section \ref{sec:heuristic} with some trial-and-error, so the directed enumeration with $N=36$ would actually finish reasonably fast and enumerate more GMA designs.
{\scriptsize
\begin{longtable}[h]{|lrrc|rrrr|}
\caption{Enumeration results: the number of isomorphism classes, real time (hours:minutes:seconds), algorithm, and partial GWP of GMA
design(s).}
\label{tab:enum} \\
\hline $\oan$ & Classes & Time & Algorithm & $A_{t+1}$ &  $A_{t+2}$
& $A_{t+3}$ & $A_{t+4}$ \\ \hline
\endfirsthead
\hline \multicolumn{8}{c}%
{\tablename\ \thetable{} -- {\rm continued \ from \ previous \
page}}\\ \hline $\oan$ & Classes & Time & Algorithm & $A_{t+1}$ &
$A_{t+2}$ & $A_{t+3}$ & $A_{t+4}$ \\ \hline
\endhead
\hline \multicolumn{8}{r}{{\rm Continued \ on \ next \ page}}\\
\hline
\endfoot
\endlastfoot
  OA(28,3,2,2) &        1 &    00:00:03 & \ref{alg:direct1}  &    0.02 &&& \\
  OA(28,4,2,2) &        3 &    00:00:01 & \ref{alg:direct1}  &    0.08 &    0.02 &&\\
  OA(28,5,2,2) &       15 &    00:00:02 & \ref{alg:direct1}  &    0.20 &    0.10 &    0.00& \\
  OA(28,6,2,2) &      320 &    00:00:36 & \ref{alg:direct1}  &    0.41 &    0.31 &    0.00 &    0.73 \\
  OA(28,7,2,2) &    12194 &    00:39:14 & \ref{alg:direct1}  &    0.71 &    0.88 &    1.55 &    0.41 \\
  OA(28,8,2,2) &    63606 &    07:16:06 & \ref{alg:direct1}  &    1.14 &    2.90 &    3.27 &    0.65 \\
  OA(28,9,2,2) &    20552 &    08:02:19 & \ref{alg:direct1}  &    1.71 &    5.51 &    6.61 &    2.45 \\
 OA(28,10,2,2) &      841 &    01:27:38 & \ref{alg:direct1}  &    2.45 &   10.49 &   11.43 &    5.06 \\
 OA(28,11,2,2) &       45 &    00:03:10 & \ref{alg:direct1}  &    3.37 &   18.82 &   14.86 &   14.69 \\
 OA(28,12,2,2) &       10 &    00:00:12 & \ref{alg:direct1}  &    4.49 &   28.22 &   25.47 &   29.39 \\
 OA(28,13,2,2) &        2 &    00:00:02 & \ref{alg:direct1}  &    5.84 &   46.43 &   32.82 &   59.43 \\
 OA(28,14,2,2) &        1 &    00:00:01 & \ref{alg:direct1}  &    7.43 &   65.00 &   52.00 &  104.00 \\
 OA(28,15,2,2) &        0 &    00:00:01 & \ref{alg:direct1}  &&&&\\
 \hline
  OA(36,3,2,2) &        1 &    00:04:41 & \ref{alg:direct2}  &    0.01 &&&\\
  OA(36,4,2,2) &        1 &    00:00:35 & \ref{alg:direct2}  &    0.05 &    0.01 &&\\
  OA(36,5,2,2) &        5 &    00:00:03 & \ref{alg:direct2}  &    0.12 &    0.06 &    0.00 &\\
  OA(36,6,2,2) &      652 &    00:04:47 & \ref{alg:direct2}  &    0.25 &    0.19 &    0.00 &    0.44 \\
  OA(36,7,2,2) &   176929 &    82:59:45 & \ref{alg:direct2}  &    0.43 &    0.43 &    1.04 &    0.35 \\
  OA(36,8,2,2) &  1320951 &  1797:40:43 & \ref{alg:direct2}  &    0.69 &    1.75 &    2.77 &    0.79 \\
  OA(36,9,2,2) &      503 &   949:40:58 & \ref{alg:direct2}  &    1.04 &    3.63 &    5.63 &    2.37 \\
 OA(36,10,2,2) &        0 &    00:07:36 & \ref{alg:direct2}  &&&&\\
 \hline
\end{longtable}}
Our new methods for directed enumerations (which could be used within ILP- or CP-based versions of the extension Step \ref{item:extend} of Algorithm \ref{alg:basic}) enumerated non-isomorphic GMA OA(28,$k$,2,2) with $k=\mbox{8-15}$ and GMA OA(36,$k$,2,2) with $k=\mbox{3-10}$ for the first time.
\section{Heuristic search for near GMA {\boldmath $\oan$}}\label{sec:heuristic}
Two search algorithms for obtaining weak and near GMA designs are defined. Their effectiveness is demonstrated against GMA designs in Table \ref{tab:bj}. Search algorithms are then used to locate weak and near GMA $\oan$ when enumerations involve too much computation.

A \emph{backward search} optimized the $\GWP$ of a randomly selected subset of $k<k_0$ columns of an $\oano$.  The subset was changed an-element-at-a-time to maximize reductions in the $\GWP$. A \emph{forward-and-backward search} starting from the OA$(N,t,s,t)$ was also needed to extend into hidden GMA $\oan$, which are not a projection of larger $\oano$.  The objective functions of ILPs from Algorithms \ref{alg:BR3idd} or \ref{alg:direct1} were randomly set to $\sum_{r=1}^{s-1}\sum_{j=1}^Nc_{rj}x_{rj}$, where sequence $c_{rj}$ for $j=1,\ldots,N$ was a random binary sequence of length $N$ with $N/s$ ones for each $r=1,\ldots,s-1$.  While feasible, the resulting ILP was used to obtain a solution and extend the design with a new factor. Whenever a design was the best to date, exhaustive search with one factor removed was performed.
\subsection{Testing heuristics against the catalog of GMA $\oan$}\label{sec:heur}
We obtained inputs $\oano$ from Neil Sloane's web catalog of OAs
and Hadamard matrices. An $N \times N$ Hadamard matrix $\HH_N$,
i.e., an orthogonal matrix with entries $\pm 1$, can be used to
obtain $N$ OA($N,N-1,2,2$) by multiplying each column of $\HH_N$
element-wise with a column of $\HH_N$, deleting the constant
column, and recoding with binary entries. (Two Hadamard matrices
are equivalent if one can be obtained from the other by permuting
columns, rows and multiplying  subsets of columns and rows by
$-1$.) Doing this once for each Hadamard inequivalent $\HH_N$ and
removing isomorphic copies produces all non-isomorphic
OA($N,N-1,2,2$). Table \ref{tab:oano} lists the number of $\oano$
we tried. All nonequivalent Hadamard matrices of order $24$ or
$28$ and the construction described above were used to produce all
non-isomorphic OA($N,N-1,2,2$) with $N=24$ and $28$.

\begin{table}[h!]\scriptsize
\caption{Number of input $\oano$ tested with backward search. An asterisk indicates that a set of all
non-isomorphic saturated designs was tested.} \label{tab:oano}
\centering \vspace{.1in}
\begin{tabular}{|l|cr|lr|r|}\hline
&\multicolumn{2}{c|}{OA$(N,N-1,2,2)$} &&&\\
$(N,k,s)$ & $\HH_N$ & Inputs & $\oano$ & Inputs & Total \\ \hline
$(24,k,2)$  & $\HH_{24}$               &   *130 &               &   &   130 \\
$(28,k,2)$  & $\HH_{28}$               & *7,570 &               &   & 7,570 \\
$(32,k,2)$  & $\HH_{16} \otimes \HH_2$ &      5 &               &   &     5 \\
$(36,k,2)$  & $\HH_{36}$               &      2 &               &   &     2 \\
$(40,k,2)$  & $\HH_{20} \otimes \HH_2$ &      3 & OA(40,20,2,3) & 1 &     4 \\
$(48,k,2)$  & $\HH_{24}\otimes \HH_2$  &    130 & OA(48,24,2,3) & 1 &   131 \\
$(81,k,3)$  &                          &        & OA(81,40,3,2) & 1 &     1 \\
$(160,k,2)$ & $\HH_{20}\otimes \HH_8$  &      2 & OA(160,80,2,3)& 1 &     3 \\ \hline
\end{tabular}
\end{table}

Our experience is that backward search quickly finds an optimum projection, even though it does not exhaustively consider all
$\binom{k_0}{k}$ projections of an $\oano$ onto $k$ columns. For example, backward search was run for each $k=3,\ldots,23$ and each OA$(24,23,2,2)$ with 100 random projections; this took 2 minutes and produced a GMA OA($24,k,2,2$) for each $k=3,\ldots,23$, where the catalog of the GWPs of GMA OA$(24,k,2,2)$ was obtained from the  OA$(24,k,2,2)$ enumeration of Schoen, Eendebak, and Nguyen~\cite{Schoen}. We used backward search and found a GMA design in 84 of the 99 cases $\oan$ listed in Table \ref{tab:bj} in 2 hours; the calculation was 100 starting random projections for each combination $(N,k,s)$ in Table \ref{tab:bj} and input $\oano$ from Table \ref{tab:oano}.

The ``hidden" GMA OA$(28,k,2,2)$ we found with $k=10$, 11, 12 are
not a projection of any OA$(28,27,2,2)$. However, if most
non-hidden GMA $\oan$ appear as projections of each
OA($N,f(N,s,t),s,t$) then backward search can efficiently find
them given an OA($N,f(N,s,t),s,t$). Table \ref{tab:extendGMA}
lists the $99-84=15$ non-located GMA $\oan$ and whether they
are a projection of a potential input design to backward
search.

\begin{table}[h!]\scriptsize
\caption{The subset of $\oan$ from Table \ref{tab:bj} where
backward search did not find a known GMA design, the number of
isomorphism classes of GMA $\oan$, and whether or not a GMA $\oan$
extends to a larger OA (determined by Algorithm
\ref{alg:basic}).  Unknown cases are indicated by question marks.
} \label{tab:extendGMA} \centering \vspace{.1in}
\begin{tabular}{|lc|cc|}\hline
$\oan$          & GMA design(s)           & Larger OA(s) & Extends \\ \hline
OA(28,6,2,2)    & 1                       & OA(28,27,2,2) & ? \\
OA(28,10,2,2)   & 2                       & OA(28,27,2,2) & no \\
OA(28,11,2,2)   & 6                       & OA(28,27,2,2) & no \\
OA(28,12,2,2)   & 2                       & OA(28,27,2,2) & no \\ \hline
OA(36,7,2,2)    & 1                       & OA(36,35,2,2) & ? \\
OA(36,8,2,2)    & 21,562                  & OA(36,35,2,2) & ? \\
OA(36,9,2,2)    & 503                     & OA(36,35,2,2) & ? \\ \hline
OA(40,$k$,2,3)  & 1 for each $k=7,8,9,10$ & OA(40,20,2,3), OA(40,39,2,2) & no, ? \\ \hline
OA(48,13,2,3)   & 1                       & OA(48,24,2,3), OA(48,47,2,2) & no, ? \\ \hline
OA(160,$k$,2,4) & 1 for each $k=7,8,9$    & OA(160,80,2,3), OA(160,159,2,2) & ?, ? \\ \hline
\end{tabular}
\end{table}

Forward-and-backward searches were performed for each $(N,s,t)$
from Section \ref{sec:results} with the ILP listed in Table \ref{tab:forwardGMA} until 1 week of computation or 1
million random objective functions. The results in Table
\ref{tab:forwardGMA} show that 78 of the 99 GMA $\oan$ listed in
Table \ref{tab:bj} were found. Forward-and-backward search found 8
of the hidden or potentially hidden GMA $\oan$ listed in Table
\ref{tab:extendGMA}. The speed of forward-and-backward search
(measured by Time$/$Current) degrades with increased $N$, but
increases with increased $t$ or use of the directed enumeration
constraints. Note that $k_{\max}=8<10=f(81,3,3)$, but for the
other $(N,s,t)$, except the $N=36$ case, the lower bounds
$k_{\max}$ all achieve the largest possible number of factors when
compared with Table \ref{tab:enum}. With $(N,s,t)=(36,2,2)$,
$k_{\max}=16>9$ (where 9 is the largest $k$ of a GMA
OA$(36,k,2,2)$ in Table \ref{tab:enum}) because Recursion
(\ref{eqn:recursive}) was not used here. Any captured
OA(36,$k$,2,2) is weak GMA because the directed enumeration constraints
of Algorithm \ref{alg:direct1} minimize $A_3$.

\begin{table}[htb!]\scriptsize
\caption{Forward-and-backward search results for the listed
$(N,s,t)$ and altered ILP. Outputs are the subset of $(N,k,s,t)$
listed in Table \ref{tab:bj} where a known GMA design was found, a
lower bound $k_{\max}$ for $f(N,s,t)$, and the approximate time
(hours:minutes) and search iteration when a better $\oan$ was last
found.} \label{tab:forwardGMA} \centering \vspace{.1in}
\begin{tabular}{|lc|crrr|}\hline
\multicolumn{2}{|c|}{Inputs} & \multicolumn{4}{c|}{Outputs} \\
$(N,s,t)$ & ILP & $k$ with GMA $\oan$ & $k_{\max}$ & Time & Iteration \\ \hline
(24,2,2)  & Algorithm \ref{alg:BR3idd}     & 3-23            & 23 & 12:00 & 60,000 \\
(28,2,2)  & Algorithm \ref{alg:direct1} & 3-5, 8, 10-14   & 14 &  3:00 & 30,000 \\
(32,2,3)  & Algorithm \ref{alg:BR3idd}     & 5-16            & 16 &  0:10 &  3,000 \\
(36,2,3)  & Algorithm \ref{alg:direct1} & 3-5             & 16 & 18:00 & 70,000 \\
(40,2,3)  & Algorithm \ref{alg:BR3idd}     & 5-20            & 20 &  5:00 & 30,000 \\
(48,2,3)  & Algorithm \ref{alg:BR3idd}     & 6, 7, 12, 15-24 & 24 & 36:00 & 50,000 \\
(81,3,3)  & Algorithm \ref{alg:BR3idd}     & 5-7             &  8 &  0:01 &     50 \\
(160,2,4) & Algorithm \ref{alg:BR3idd}     & 7               &  9 &  1:00 &     80 \\
\hline
\end{tabular}
\end{table}
\subsection{Finding good designs by using heuristics}
Table \ref{tab:weakN36} has partial GWPs of the best OA(36,$k$,2,2) obtained from the heuristic search algorithms. The forward-and-backward searches from Section \ref{sec:heur} were used directly, but $R=1,000$ backward searches were run for each $k=3,\ldots,19$ on each of the 2 input OA$(36,35,2,2)$ we tried. These more extensive backward searches all finished within 6 minutes.  Designs with $k \le 6$ were not provided because comparison with Table \ref{tab:bj} showed that these designs were GMA. However, by Table \ref{tab:enum}, when $7 \le k \le 9$, the search algorithms found weak GMA designs which are not GMA. Designs with $10\leq k \leq 18$ are also weak GMA because each achieves the lowest possible value of $A_3 = \binom{k}{3}4^2/36^2$. It is unknown to us which if any of our best found OA(36,$k$,2,2) with $k>9$ are GMA.

\begin{table}[htb!]\scriptsize
\caption{Partial GWPs of weak or near GMA OA$(36,k,2,2)$ from heuristic searches.} \label{tab:weakN36} \centering \vspace{.1in}
\begin{tabular}{|lc|rrrr|}\hline
$\oan$ & Search method & $A_3$ & $A_4$ & $A_5$ & $A_6$ \\ \hline
  OA(36,7,2,2) &   backward  &    0.43 &    0.73 &    0.94 &    0.44 \\
  OA(36,8,2,2) & forward-and-backward  &    0.69 &    1.85 &    2.57 &    0.89 \\
  OA(36,9,2,2) &  forward-and-backward  &    1.04 &    3.93 &    5.38 &    2.07 \\ \hline
 OA(36,10,2,2) & forward-and-backward  &    1.48 &    7.33 &    9.93 &    4.69 \\
 OA(36,11,2,2) &  forward-and-backward  &    2.04 &   13.46 &   13.33 &   12.54 \\
 OA(36,12,2,2) &  forward-and-backward  &    2.72 &   22.70 &   16.79 &   28.54 \\
 OA(36,13,2,2) & forward-and-backward  &    3.53 &   33.91 &   24.44 &   54.72 \\
 OA(36,14,2,2) &  forward-and-backward  &    4.49 &   47.81 &   37.78 &   94.72 \\
 OA(36,15,2,2) &  forward-and-backward  &    5.62 &   65.44 &   56.64 &  156.79 \\
 OA(36,16,2,2) & forward-and-backward  &    6.91 &   87.26 &   82.57 &  250.86 \\
 OA(36,17,2,2) &  backward  &    8.40 &  123.41 &  100.94 &  423.11 \\
 OA(36,18,2,2) &  backward   &   10.07 &  158.67 &  141.04 &  634.67 \\ \hline
 OA(36,19,2,2) &  backward  &   17.30 &  168.74 &  261.33 &  775.70 \\\hline
\end{tabular}
\end{table}

Backward searches for each of the 7,570 non-isomorphic OA(28,27,2,2) and $k=15,\ldots,26$ were run with $R=100$ and finished within 10 hours. Forward-and-backward search was also rerun with $(N,s,t)=(28,2,2)$ as explained in Section \ref{sec:heur}, except that Algorithm \ref{alg:BR3idd} ILPs were used instead to attain larger $k_{\max}$. This job ran for the full week and outputted $k_{\max}=27=f(28,2,2)$, but better designs were not found after 1 day. Table \ref{tab:heurN28} has partial GWPs of the best OA(28,$k$,2,2) with $k=\mbox{14-27}$; the GMA and weak GMA status of designs with $15 \le k \le 26$ is unknown. The designs with the truncated GWPs in Tables \ref{tab:weakN36} and \ref{tab:heurN28} are available upon request.

\begin{table}[htb!]\scriptsize
\caption{Partial GWPs of near GMA OA$(28,k,2,2)$ from heuristic searches.}
\label{tab:heurN28} \centering \vspace{.1in}
\begin{tabular}{|lc|rrrr|}\hline
$\oan$ & Search method & $A_3$ & $A_4$ & $A_5$ & $A_6$ \\ \hline
 OA(28,15,2,2) &  forward-and-backward  &   12.71 &   72.43 &   97.71 &  156.00 \\
 OA(28,16,2,2) & forward-and-backward   &   17.31 &   87.43 &  145.63 &  263.18 \\
 OA(28,17,2,2) &  forward-and-backward  &   23.18 &  105.06 &  218.20 &  407.84 \\
 OA(28,18,2,2) & forward-and-backward  &   29.39 &  129.06 &  306.12 &  623.02 \\
 OA(28,19,2,2) &  forward-and-backward  &   36.59 &  159.10 &  412.73 &  924.90 \\
 OA(28,20,2,2) &  forward-and-backward   &   43.84 &  196.67 &  549.22 & 1330.29 \\
 OA(28,21,2,2) &  forward-and-backward   &   51.63 &  241.82 &  718.04 & 1870.37 \\
 OA(28,22,2,2) & backward,  forward-and-backward  &   60.82 &  293.45 &  922.20 & 2588.41 \\
 OA(28,23,2,2) &  backward,  forward-and-backward  &   70.43 &  354.59 & 1174.12 & 3507.67 \\
 OA(28,24,2,2) & backward,  forward-and-backward  &   80.49 &  425.51 & 1483.10 & 4676.90 \\
 OA(28,25,2,2) & backward,  forward-and-backward   &   92.00 &  506.00 & 1848.00 & 6160.00 \\
 OA(28,26,2,2) & backward,  forward-and-backward   &  104.00 &  598.00 & 2288.00 & 8008.00 \\ \hline
 OA(28,27,2,2) &  $\HH_{28}$  &  117.00 &  702.00 & 2808.00 & 10296.00 \\ \hline
\end{tabular}
\end{table}
\newpage\section{Conclusion}\label{sec:sum}
The fundamental problem with enumerating $\oan$ is that the number of ILPs that need to be solved grows exponentially with $N$. Our method of directed enumeration reduced this number and helped find more GMA designs; see the 28 and 36 run cases in Table \ref{tab:enum}. The heuristic searches in Section \ref{sec:heuristic} quickly found GMA $\oan$, matching 92 of the 99 distance distributions of GMA designs given in Table \ref{tab:bj}. Future research for finding GMA $\oan$ will involve developing CP with isomorphism rejection algorithms based on adding rows that satisfy a set of non-linear constraints.
\section*{Acknowledgements}
The authors are indebted to Professor~Francois~Margot for providing ILP solvers and answering questions. The authors also owe much thanks to Professor Brendan McKay for answering questions regarding his program \texttt{nauty}.

This research was supported by the AFOSR grants F1ATA06334J001, F1ATA03039J001, and by an allocation of computing time from the Ohio Supercomputer Center.  The views expressed in this article are those of the authors and do not reflect the official policy or position of the United States Air Force, Department of Defense, or the U.S.~Government.

\begin{appendix}{}
\section{Theoretical justifications}\label{sec:proofs}
\subsection{Proof of Lemma \ref{lem:recursion}}
By Lemma \ref{lem:divisibility}, it suffices to show $A_{t+r} (\Y(-i))\leq  A_{t+r}
(\Y)- (t+r) A_{t+r} (\Y)/k$ for some factor $i$. Each $J_{t+r}(l)$
involves $t+r$ factors, so on average each factor contributes
$(t+r)A_{t+r}/k$ to $A_{t+r}$. Then there must exist a factor that
contributes at least  $(t+r)A_{t+r}/k$ to $A_{t+r}$. Let this
factor be factor $i$.  Then $A_{t+r} (\Y(-i))\leq  A_{t+r} (\Y)-
(t+r) A_{t+r} (\Y)/k$.
\section{Distance distributions of GMA designs}\label{sec:cat}\scriptsize

\begin{longtable}[h]{|l|l|}
\caption{Distance distributions of GMA designs.} \label{tab:bj} \\
\hline $\oan$ & $B_0,~B_1,~\ldots,~B_k$  \\ \hline
\endfirsthead
\hline \multicolumn{2}{c}%
{\tablename\ \thetable{} -- {\rm continued \ from \ previous \
page}}\\ \hline $\oan$ & $B_0,~B_1,\ldots,~B_k$  \\ \hline
\endhead
\hline \multicolumn{2}{r}{{\rm Continued \ on \ next \ page}}\\
\hline
\endfoot
\endlastfoot
  OA(24,3,2,2) & 3, 9, 9, 3 \\
  OA(24,4,2,2) & 1.667, 5.333, 10, 5.333, 1.667 \\
  OA(24,5,2,2) & 1.167, 2.500, 8.333, 8.333, 2.500, 1.167 \\
  OA(24,6,2,2) & 1, 1, 5, 10, 5, 1, 1 \\
  OA(24,7,2,2) & 1, 0.167, 2.500, 8.333, 8.333, 2.500, 0.167, 1 \\
  OA(24,8,2,2) & 1, 0, 0.667, 5.333, 10, 5.333, 0.667, 0, 1 \\
  OA(24,9,2,2) & 1, 0, 0, 2, 9, 9, 2, 0, 0, 1 \\
 OA(24,10,2,2) & 1, 0, 0, 0, 5, 12, 5, 0, 0, 0, 1 \\
 OA(24,11,2,2) & 1, 0, 0, 0, 0, 11, 11, 0, 0, 0, 0, 1 \\
 OA(24,12,2,2) & 1, 0, 0, 0, 0, 0, 22, 0, 0, 0, 0, 0, 1 \\
 OA(24,13,2,2) & 1, 0, 0, 0, 0, 0, 10, 12, 0, 0, 0, 0, 1, 0 \\
 OA(24,14,2,2) & 1, 0, 0, 0, 0, 0, 4, 12, 6, 0, 0, 0, 1, 0, 0 \\
 OA(24,15,2,2) & 1, 0, 0, 0, 0, 0, 1.333, 8, 10, 2.667, 0, 0, 1, 0, 0, 0 \\
 OA(24,16,2,2) & 1, 0, 0, 0, 0, 0, 0.333, 4, 10, 6.667, 1, 0, 1, 0, 0, 0, 0 \\
 OA(24,17,2,2) & 1, 0, 0, 0, 0, 0, 0, 1.333, 7.667, 9.333, 2.667, 1.333, 0.667, 0, 0, 0, 0, 0 \\
 OA(24,18,2,2) & 1, 0, 0, 0, 0, 0, 0, 0, 3, 12, 6, 0, 2, 0, 0, 0, 0, 0, 0 \\
 OA(24,19,2,2) & 1, 0, 0, 0, 0, 0, 0, 0, 0, 9, 9, 3, 2, 0, 0, 0, 0, 0, 0, 0 \\
 OA(24,20,2,2) & 1, 0, 0, 0, 0, 0, 0, 0, 0, 0, 18, 0, 5, 0, 0, 0, 0, 0, 0, 0, 0 \\
 OA(24,21,2,2) & 1, 0, 0, 0, 0, 0, 0, 0, 0, 0, 6, 12, 5, 0, 0, 0, 0, 0, 0, 0, 0, 0 \\
 OA(24,22,2,2) & 1, 0, 0, 0, 0, 0, 0, 0, 0, 0, 0, 12, 11, 0, 0, 0, 0, 0, 0, 0, 0, 0, 0 \\
 OA(24,23,2,2) & 1, 0, 0, 0, 0, 0, 0, 0, 0, 0, 0, 0, 23, 0, 0, 0, 0, 0, 0, 0, 0, 0, 0, 0 \\
 \hline
  OA(28,3,2,2) & 3.571, 10.286, 10.714, 3.429 \\
  OA(28,4,2,2) & 1.929, 6.571, 10.714, 7.143, 1.643 \\
  OA(28,5,2,2) & 1.143, 3.929, 8.571, 9.286, 4.286, 0.786 \\
  OA(28,6,2,2) & 1.071, 0.429, 10.714, 2.857, 11.786, 0.429, 0.714 \\
  OA(28,7,2,2) & 1, 0, 4.786, 8.857, 6.714, 4.857, 1.786, 0 \\
  OA(28,8,2,2) & 1, 0, 1.429, 7.429, 9, 5.143, 2.857, 1.143, 0 \\
  OA(28,9,2,2) & 1, 0, 0, 4.500, 9.643, 6.429, 3.857, 1.929, 0.643, 0 \\
 OA(28,10,2,2) & 1, 0, 0, 1.143, 7.429, 10, 4.429, 2.286, 1.143, 0.571, 0 \\
 OA(28,11,2,2) & 1, 0, 0, 0.214, 3.643, 9.429, 7.929, 4, 0.929, 0, 0.786, 0.071 \\
 OA(28,12,2,2) & 1, 0, 0, 0, 0.643, 7.714, 9.857, 5.143, 2.786, 0, 0, 0.857, 0 \\
 OA(28,13,2,2) & 1, 0.071, 0, 0, 0, 2.571, 9.143, 11.286, 3, 0, 0, 0, 0.857, 0.071 \\
 OA(28,14,2,2) & 1, 0.071, 0, 0, 0, 0, 5.571, 13, 7.429, 0, 0, 0, 0, 0.929, 0 \\
 \hline
  OA(32,4,2,3) & 2, 8, 12, 8, 2 \\
  OA(32,5,2,3) & 1, 5, 10, 10, 5, 1 \\
  OA(32,6,2,3) & 1, 0, 15, 0, 15, 0, 1 \\
  OA(32,7,2,3) & 1, 0, 5, 12, 7, 4, 3, 0 \\
  OA(32,8,2,3) & 1, 0, 1, 10, 11, 4, 3, 2, 0 \\
  OA(32,9,2,3) & 1, 0, 0, 4, 14, 8, 0, 4, 1, 0 \\
 OA(32,10,2,3) & 1, 0, 0, 0, 10, 16, 0, 0, 5, 0, 0 \\
 OA(32,11,2,3) & 1, 0, 0, 0, 5, 10, 10, 5, 0, 0, 0, 1 \\
 OA(32,12,2,3) & 1, 0, 0, 0, 1, 8, 12, 8, 1, 0, 0, 0, 1 \\
 OA(32,13,2,3) & 1, 0, 0, 0, 0, 3, 12, 12, 3, 0, 0, 0, 0, 1 \\
 OA(32,14,2,3) & 1, 0, 0, 0, 0, 0, 7, 16, 7, 0, 0, 0, 0, 0, 1 \\
 OA(32,15,2,3) & 1, 0, 0, 0, 0, 0, 0, 15, 15, 0, 0, 0, 0, 0, 0, 1 \\
 OA(32,16,2,3) & 1, 0, 0, 0, 0, 0, 0, 0, 30, 0, 0, 0, 0, 0, 0, 0, 1 \\
 \hline
  OA(36,3,2,2) & 4.556, 13.333, 13.667, 4.444 \\
  OA(36,4,2,2) & 2.389, 8.667, 13.667, 9.111, 2.167 \\
  OA(36,5,2,2) & 1.333, 5.278, 11.111, 11.667, 5.556, 1.056 \\
  OA(36,6,2,2) & 1.056, 1.667, 11.667, 6.667, 12.500, 1.667, 0.778 \\
  OA(36,7,2,2) & 1, 0, 8.167, 7.778, 11.667, 4.667, 2.722, 0 \\
  OA(36,8,2,2) & 1, 0, 2.667, 9.778, 10.333, 6.667, 4.222, 1.333, 0 \\
  OA(36,9,2,2) & 1, 0, 0.333, 7, 11, 8, 5, 3, 0.667, 0 \\
 \hline
  OA(40,4,2,3) & 2.600, 9.600, 15.600, 9.600, 2.600 \\
  OA(40,5,2,3) & 1.500, 5.500, 13, 13, 5.500, 1.500 \\
  OA(40,6,2,3) & 1, 3, 9, 14, 9, 3, 1 \\
  OA(40,7,2,3) & 1, 0.500, 7.500, 11, 11, 7.500, 0.500, 1 \\
  OA(40,8,2,3) & 1.100, 0, 2.400, 11.200, 13, 6.400, 4, 1.600, 0.300 \\
  OA(40,9,2,3) & 1.100, 0, 0, 7.200, 14.400, 9, 3.600, 3.600, 0.900, 0.200 \\
 OA(40,10,2,3) & 1.100, 0, 0, 0, 18, 7.200, 9, 0, 4.500, 0, 0.200 \\
 OA(40,11,2,3) & 1, 0, 0, 1.200, 6.400, 11.400, 11.400, 6.400, 1.200, 0, 0, 1 \\
 OA(40,12,2,3) & 1, 0, 0, 0, 3.600, 9.600, 11.600, 9.600, 3.600, 0, 0, 0, 1 \\
 OA(40,13,2,3) & 1, 0, 0, 0, 0.900, 6.300, 11.800, 11.800, 6.300, 0.900, 0, 0, 0, 1 \\
 OA(40,14,2,3) & 1, 0, 0, 0, 0, 3, 9, 14, 9, 3, 0, 0, 0, 0, 1 \\
 OA(40,15,2,3) & 1, 0, 0, 0, 0, 0.500, 5.500, 13, 13, 5.500, 0.500, 0, 0, 0, 0, 1 \\
 OA(40,16,2,3) & 1, 0, 0, 0, 0, 0, 1.600, 9.600, 15.600, 9.600, 1.600, 0, 0, 0, 0, 0, 1 \\
 OA(40,17,2,3) & 1, 0, 0, 0, 0, 0, 0, 4, 15, 15, 4, 0, 0, 0, 0, 0, 0, 1 \\
 OA(40,18,2,3) & 1, 0, 0, 0, 0, 0, 0, 0, 9, 20, 9, 0, 0, 0, 0, 0, 0, 0, 1 \\
 OA(40,19,2,3) & 1, 0, 0, 0, 0, 0, 0, 0, 0, 19, 19, 0, 0, 0, 0, 0, 0, 0, 0, 1 \\
 OA(40,20,2,3) & 1, 0, 0, 0, 0, 0, 0, 0, 0, 0, 38, 0, 0, 0, 0, 0, 0, 0, 0, 0, 1 \\
 \hline
  OA(48,4,2,3) & 3, 12, 18, 12, 3 \\
  OA(48,5,2,3) & 1.667, 6.667, 16.667, 13.333, 8.333, 1.333 \\
  OA(48,6,2,3) & 1.167, 3, 12.833, 15.333, 9.500, 5.667, 0.500 \\
  OA(48,7,2,3) & 1, 1, 8, 16, 11, 7, 4, 0 \\
  OA(48,8,2,3) & 1, 0, 4, 14, 13, 8, 6, 2, 0 \\
  OA(48,9,2,3) & 1, 0, 0.667, 10, 15.333, 9.333, 6, 4.667, 1, 0 \\
 OA(48,10,2,3) & 1, 0, 0, 4, 15, 13, 6, 6, 2, 1, 0 \\
 OA(48,11,2,3) & 1, 0, 0, 0.333, 10.333, 17.667, 6.333, 5.667, 5.333, 0.333, 1, 0 \\
 OA(48,12,2,3) & 1, 0, 0, 0, 4, 13.333, 17.333, 5.333, 1.667, 2.667, 2.667, 0, 0 \\
 OA(48,13,2,3) & 1, 0, 0, 0, 0, 9, 22, 6, 0, 9, 0, 0, 1, 0 \\
 OA(48,14,2,3) & 1, 0, 0, 0, 1, 0, 15, 24, 0, 0, 5, 0, 2, 0, 0 \\
 OA(48,15,2,3) & 1, 0, 0, 0, 0, 1.667, 9, 12.333, 12.333, 9, 1.667, 0, 0, 0, 0, 1 \\
 OA(48,16,2,3) & 1, 0, 0, 0, 0, 0, 5, 12, 12, 12, 5, 0, 0, 0, 0, 0, 1 \\
 OA(48,17,2,3) & 1, 0, 0, 0, 0, 0, 1, 9, 13, 13, 9, 1, 0, 0, 0, 0, 0, 1 \\
 OA(48,18,2,3) & 1, 0, 0, 0, 0, 0, 0, 4, 11, 16, 11, 4, 0, 0, 0, 0, 0, 0, 1 \\
 OA(48,19,2,3) & 1, 0, 0, 0, 0, 0, 0, 0.667, 7, 15.333, 15.333, 7, 0.667, 0, 0, 0, 0, 0, 0, 1 \\
 OA(48,20,2,3) & 1, 0, 0, 0, 0, 0, 0, 0, 2, 12, 18, 12, 2, 0, 0, 0, 0, 0, 0, 0, 1 \\
 OA(48,21,2,3) & 1, 0, 0, 0, 0, 0, 0, 0, 0, 5, 18, 18, 5, 0, 0, 0, 0, 0, 0, 0, 0, 1 \\
 OA(48,22,2,3) & 1, 0, 0, 0, 0, 0, 0, 0, 0, 0, 11, 24, 11, 0, 0, 0, 0, 0, 0, 0, 0, 0, 1 \\
 OA(48,23,2,3) & 1, 0, 0, 0, 0, 0, 0, 0, 0, 0, 0, 23, 23, 0, 0, 0, 0, 0, 0, 0, 0, 0, 0, 1 \\
 OA(48,24,2,3) & 1, 0, 0, 0, 0, 0, 0, 0, 0, 0, 0, 0, 46, 0, 0, 0, 0, 0, 0, 0, 0, 0, 0, 0, 1 \\
 \hline
  OA(81,4,3,3) & 1, 8, 24, 32, 16 \\
  OA(81,5,3,3) & 1, 0, 20, 20, 30, 10 \\
  OA(81,6,3,3) & 1, 0, 4, 24, 24, 20, 8 \\
  OA(81,7,3,3) & 1, 0, 0, 10, 30, 18, 16, 6 \\
  OA(81,8,3,3) & 1, 0, 0, 0, 20, 32, 8, 16, 4 \\
  OA(81,9,3,3) & 1, 0, 0, 0, 0, 36, 24, 0, 18, 2 \\
 OA(81,10,3,3) & 1, 0, 0, 0, 0, 0, 60, 0, 0, 20, 0 \\
 \hline
  OA(96,5,2,4) & 3, 15, 30, 30, 15, 3 \\
  OA(96,6,2,4) & 1.667, 8, 25, 26.667, 25, 8, 1.667 \\
  OA(96,7,2,4) & 1.167, 3.500, 18.167, 25.833, 24.167, 17.833, 4.500, 0.833 \\
 \hline
 OA(112,5,2,4) & 3.571, 17.143, 35.714, 34.286, 17.857, 3.429 \\
 OA(112,6,2,4) & 2, 9.429, 27.857, 34.286, 25.714, 11.143, 1.571 \\
 \hline
 OA(128,5,2,4) & 4, 20, 40, 40, 20, 4 \\
 OA(128,6,2,4) & 2, 12, 30, 40, 30, 12, 2 \\
 OA(128,7,2,4) & 1, 7, 21, 35, 35, 21, 7, 1 \\
 OA(128,8,2,4) & 1, 0, 28, 0, 70, 0, 28, 0, 1 \\
 OA(128,9,2,4) & 1, 0, 9, 27, 27, 27, 27, 9, 0, 1 \\
OA(128,10,2,4) & 1, 0, 3, 19, 29, 27, 25, 17, 6, 1, 0 \\
OA(128,11,2,4) & 1, 0, 0, 12, 26, 28, 24, 20, 13, 4, 0, 0 \\
OA(128,12,2,4) & 1, 0, 0, 3, 21, 35, 19, 17, 22, 9, 1, 0, 0 \\
OA(128,13,2,4) & 1, 0, 0, 0, 10, 36, 28, 8, 21, 20, 4, 0, 0, 0 \\
OA(128,14,2,4) & 1, 0, 0, 0, 0, 28, 42, 8, 7, 28, 14, 0, 0, 0, 0 \\
OA(128,15,2,4) & 1, 0, 0, 0, 0, 0, 70, 0, 15, 0, 42, 0, 0, 0, 0, 0 \\
 \hline
 OA(144,5,2,4) & 4.556, 22.222, 45.556, 44.444, 22.778, 4.444 \\
 OA(144,6,2,4) & 2.444, 12.667, 35, 44.444, 33.333, 14, 2.111 \\
 OA(144,7,2,4) & 1.528, 6.417, 25.083, 39.861, 37.917, 23.917, 8.361, 0.917 \\
 OA(144,8,2,4) & 1.208, 2.556, 16.722, 33.444, 37.917, 30.333, 16.722, 4.778, 0.319 \\
 \hline
 OA(160,5,2,4) & 5, 25, 50, 50, 25, 5 \\
 OA(160,6,2,4) & 2.600, 14.400, 39, 48, 39, 14.400, 2.600 \\
 OA(160,7,2,4) & 1.600, 7, 29.400, 42, 42, 29.400, 7, 1.600 \\
 OA(160,8,2,4) & 1.100, 3.300, 18.900, 36.700, 40.500, 35.900, 18.700, 4.100, 0.800 \\
 OA(160,9,2,4) & 1, 0.900, 11.400, 30, 37, 37, 28.600, 11.600, 2, 0.500 \\
 \hline
 OA(176,5,2,4) & 5.545, 27.273, 55.455, 54.545, 27.727, 5.455 \\
 OA(176,6,2,4) & 2.909, 15.818, 42.273, 54.545, 40.909, 16.909, 2.636 \\
 OA(176,7,2,4) & 1.705, 8.432, 30.068, 48.523, 46.932, 29.114, 10.023, 1.205 \\
 OA(176,8,2,4) & 1.216, 3.909, 20.045, 40.091, 46.932, 37.545, 20.045, 5.727, 0.489 \\
 \hline
\end{longtable}
\end{appendix}
\end{document}